\documentclass{PoS}

\usepackage{ctable}
\usepackage{pdflscape}
\usepackage{amsmath}
\title{Assessment of a statistical approach that facilitates the constraint of pulsar geometry via dualband light curve fitting}

\ShortTitle{A statistical approach to constraining pulsar geometry via dualband LC fitting}

\author{\speaker{Mechiel C. Bezuidenhout}, Christo Venter, and Albertus S. Seyffert\\
        Centre for Space Research, North-West University, 11 Hoffman Street, Potchefstroom, 2531,
South Africa\\
        E-mail: \email{bezmc93@gmail.com}}
        
\author{Alice K. Harding\\
        Astrophysics Science Division, NASA Goddard Space Flight Center, 8800 Greenbelt Rd,
Greenbelt, MD 20771, United States\\}

\abstract{The Large Area Telescope aboard the \textit{Fermi} spacecraft has detected more than 200 $\gamma$-ray
pulsars since its launch in 2008. By concurrently fitting standard geometric model light curves onto
\textit{Fermi} and radio data, researchers have constrained the inclination and observer angles of a number
of pulsars. At first this was done by comparing observed and modelled light curves by eye, and later
via statistical approaches. We fit modelled light curves of 16 pulsars to radio and $\gamma$-ray data by optimising a custom test statistic that we have
developed for combining light curves across the two wavebands, taking their disparate errors into account. We present geometrical constraints found using this process, and
compare them with results found by eye or using other statistical methods.}

\FullConference{5th Annual Conference on High Energy Astrophysics in Southern Africa\\
		4-6 October, 2017\\
		University of the Witwatersrand (Wits), South Africa}

\begin{document}

\section{Introduction}
 Pulsar observations have traditionally been dominated by data from the radio wavelength range, with sources in this band totalling 2 627 at the time of writing, according to the ATNF Pulsar Catalogue \cite{manchester2005}. This is in comparison with the fewer than ten sources that had been detected in the $\gamma$-ray domain prior to 2008 \cite{thompson2001}. The launch of the Large Area Telescope (LAT) aboard the \textit{Fermi} satellite that year marked the deployment of the first instrument sensitive enough to detect a significant number or $\gamma$-ray photons from pulsars. To date, it has discovered more than 205 new $\gamma$-ray pulsars at high sensitivity and 
   resolution, ushering in what's been called a $\gamma$-ray pulsar revolution \cite{caraveo2014}. 
   
Multi-wavelength pulsar studies can now be done that aid our understanding of 
pulsar emission in several wavelength domains. In particular, using some goodness-of-fit test to compare observed and modelled pulsar light curves (LCs) in the radio and $\gamma$-ray bands, one can infer a pulsar's 
most likely physical configuration, in terms of its tilt angle ($\alpha$) and the observer angle ($\zeta$) (see e.g. \cite{venter2009}). Unfortunately, the available $\gamma$-ray and radio geometric models often produce very different $\alpha$ and $\zeta$ constraints
when fit separately to observations. 

Various attempts have been made to find "compromise" ($\alpha,\zeta$) constraints, i.e.\ pairs of these parameters that lead to model LCs that adequately match observations in both bands. This endeavour has, however, been complicated by the disparity between the errors characterising the available radio and $\gamma$-ray pulsar observations. Due to this error discrepancy, any goodness-of-fit test that is dependent on the observational errors, such as Pearson's $\chi^2$ test, will deliver very different values of the test statistic in the radio and $\gamma$-ray wavebands. Adding the test statistics obtained for individual fits and then minimising the result typically leads to a good fit in one waveband and a relatively poor fit in the other. 

Previous studies have applied Pearson's $\chi^2$ goodness-of-fit test to compare modelled LCs to radio and $\gamma$-ray observations of millisecond pulsars (MSPs) \cite{pierbattista2015} and canonical pulsars \cite{johnson2014}. These studies attempted to circumvent the error-disparity problem by artificially inflating the relative uncertainties on the observed radio data to match those of the $\gamma$-ray observations before addition of the respective test statistics of the radio-only and $\gamma$-only LC fits. This produced LC fits in both wavebands that are invariably qualitatively better than those obtained without taking error disparity into account. Compromising the radio data is, however, not optimal for achieving the best possible LC fits, and deliver formally bad fits, the minimum value of the test statistic being orders of magnitude larger than the number of degrees of freedom. Furthermore, it has been suggested to scale the dynamic range of Pearson's $\chi^2$ test statistic to match in the radio and $\gamma$-ray bands before addition \cite{seyffert2016}. This method was recently applied to a sample of 11 MSPs and canonical pulsars, with mixed results \cite{bezuidenhout2017}.

Recently, a bespoke test statistic called the Scaled-Flux Normalised $\chi^2$ test statistic has been put forward as an alternative to Pearson's $\chi^2$ test (Seyffert et al., 2018; in preparation). This test statistic partially negates the influence of significantly differing data errors, making it comparable between wavebands without the need for error inflation, scaling, or any other manipulation. In this work we apply the proposed test statistic to a sample of 16 pulsars previously fit by either of the previous studies (\cite{pierbattista2015},\cite{johnson2014}). We then compare the best-fit $(\alpha,\zeta)$\footnote{Note that pairs of constraints on $\alpha$ and $\zeta$ are reported throughout in this order and as measured in degrees.} pairs we find using the SFN test statistic to those found by the previous studies, as well as those obtained using by-eye comparison of modelled and observed LCs. For more details see Bezuidenhout et al. (2018; in preparation).

\section{LC fitting methods}
Various methods have been applied to find ($\alpha,\zeta$) pairs that lead to the best possible match between modelled and observed pulsar LCs in both the radio and $\gamma$-ray wavebands simultaneously. Here we exposit the most notable of these methods, as well as the newly proposed SFN fitting method.

\subsection{By-eye fitting}

The first approach used in this work towards finding best-fit ($\alpha$,$\zeta$) pairs is simple by-eye fitting of modelled LCs onto observed data.  For this purpose we modelled the pulsars' radio emission as being either a simple, Gaussian beam (core model) or a hollow cone (cone model) \cite{story2007}, and their $\gamma$-ray emission using either the Two-Pole Caustic (TPC) \cite{dyks2003} or Outer Gap (OG) \cite{cheng1987} geometrical models. 

For a given pulsar, $\gamma$-ray and radio LCs are modelled assuming a particular ($\alpha$,$\zeta$) combination, and then
superimposed on the pulsar's observed LCs. We make a qualitative decision as to whether or
not the experimental data are satisfactorily reproduced by this LC realisation. In making this judgement, particular attention is paid to whether or not the number of $\gamma$-ray and radio peaks in the modelled LC matches that of the observed LC, whether the peaks have similar shapes, and whether the peaks occur at similar phases. This 
process is iterated to cover all possible combinations of $\alpha$ and $\zeta$, keeping all other model parameters fixed. The best-fit ($\alpha$,$\zeta$) is specified as the centre of the region in $\alpha$-$\zeta$ space where the resultant modelled LCs are adjudged to adequately fit the observed LC, with their errors defining a box encompassing this whole region.

This fitting method is, of course, rather subjective:  applying this method twice to the
same pulsar using the same geometric model will result in two slightly differing answers. This is
especially true when there are multiple ($\alpha$,$\zeta$) pairs plausibly replicating the observed data for
a single pulsar. As such, this fitting method is not seen as a serious attempt at constraining pulsar
geometries in and of itself, but rather as a sanity check, or a basis for (qualitatively)
judging the accuracy of more rigorous methods: if a statistical approach produces an answer far
out of line with what by-eye fitting delivers, the former result may be cast into doubt. Conversely, a strong correspondence
between the results of by-eye fitting and a given statistical fit only serves to strengthen the case for the suitability of the statistical fitting method in question.

\subsection{Pearson's $\chi^2$ goodness-of-fit test}

Pearson's $\chi^2$ test statistic as applicable to pulsar LC fitting can be written as
\begin{equation}
\chi^{2} = \sum_{i=1}^{n_{\rm{bins}}} \left( \frac{E_{i} - O_{i}}{U_{i}} \right)^{2}, 
\end{equation}
where $E_{i}$, $O_{i}$, and $U_{i}$ respectively refer to the modelled (expected) intensity, observed intensity, and uncertainty on the observed intensity in the $i$th bin of $n_{\rm{bins}}$ bins of the LC \cite{pearson1900}. 

If our LC models have $n_{\rm{parameters}}$ free parameters, the minimum value of $\chi^{2}$ would ideally be approximately equal to the number of degrees of freedom, $N_{\rm{dof}} = n_{\rm{bins}}-n_{\rm{parameters}}$, specifying a good match between modelled and observed data. However, this rarely occurs for the models currently available, with the minimum test statistic often being far larger than $N_{\rm{dof}}$. This indicates that the models are still somewhat rough approximations of the real phenomena. 

Typically, finding the ($\alpha$,$\zeta$) pair that leads to the best fit of a pulsar's modelled and observed LCs entails the following: first, the $\chi^2$ test is applied to all possible ($\alpha$,$\zeta$) pairs in radio and $\gamma$ rays separately, yielding two "maps" of test statistic values on axes of $\alpha$ and $\zeta$. These maps can then be independently minimised to find the best-fit combination of $\alpha$ and $\zeta$ for each waveband. This minimum may or may not be co-located in the radio and $\gamma$-ray bands.  

The purpose of our work is to find a \textit{combined} $\gamma$-ray and radio fit, i.e. an ($\alpha$,$\zeta$) pair corresponding to an adequate simultaneous radio and $\gamma$-ray fit. The standard approach to finding such a combined fit is to add the separate radio and $\gamma$-ray $\chi^2$ maps, and then minimise the resultant. The dependence of Pearson's test statistic on observational uncertainties $U_i$ creates problems, however. In our case, the relative errors on the observed $\gamma$-ray data are much larger than they are on the radio data, so that the typical values of $\chi^{2}$ are much smaller for the LC in the $\gamma$-ray domain than for the radio domain. Simply adding the test statistics in the $\gamma$-ray and radio domains therefore creates a combined $\chi^{2}$ map that is very much radio-dominated. When using this combined fitting method it is therefore common to end up with a model fit that reproduces observed radio LC qualitatively well (although not formally), but the observed $\gamma$-ray LC not at all.  

\subsection{The Scaled-Flux Normalised $\chi^2$ test statistic} 
It is clear that the Pearson $\chi^2$ test statistic's dependence on observational data uncertainties makes it ill suited for combining fits across data sets with disparate errors. For this reason Seyffert et al. (2018; in preparation) developed an alternate test statistic which minimises the impact of such data error disparities, called the Scaled-Flux Normalised (SFN; $\chi_\phi^2$) test statistic. The SFN test statistic is defined as 

\begin{equation}\label{eq:chi2phi}
\chi_\phi^2 = \frac{\phi^2 - \chi^2}{\phi^2 - n_{\rm{dof}}},
\end{equation}    
where 
\begin{equation}\label{eq:phi2}
\phi^2 = \sum_{i}^{n_{\rm{bins}}} \left(\frac{D_i - B_i}{\epsilon_i}\right)^2
\end{equation}
is effectively the value of Pearson's $\chi^2$ test statistic for the background (off-peak) flux. The $D_i$, $B_i$, and $\epsilon_i$ are the observed flux, background flux, and error on the observed flux in the $i^{\rm{th}}$ bin, respectively. The test statistic's name refers to the fact that it represents a normalised measure of the goodness-of-fit of the "excess", non-background, flux.

It is instructive to consider the case of a very good match between modelled and observed LCs: under these conditions $\chi^2 \sim n_{\rm{dof}}$, so that the value of the test statistic will be approximately unity. On the other end of the scale, e.g. in the case where the model predicts nothing but background noise, $\chi^2 \sim \phi^2$, and thus the SFN test statistic approaches zero. A negative value (indicating a worse fit than a flat model LC) is unlikely, and a value larger than one represents a suspicious over-fit. Therefore the value of the SFN test statistic can be thought of as a normalised measure of goodness-of-fit, scaling from zero (no fit) to one (good fit). Maximising this test statistic across all possible combinations of $\alpha$ and $\zeta$ results in the best possible estimation of a given pulsar's geometry.

The most crucial facet of the SFN test statistic is that it partially negates the problem presented by comparisons between data sets with dissimilar errors. As opposed to Pearson's test statistic, which returns values that can vary greatly depending on the data errors, the SFN test statistic nearly always returns values around the range of zero to one. This means that SFN test statistic values are much more consistently comparable between data sets, regardless of uncertainties, than is the case with Pearson's test statistic, rendering any sort of error inflation or scaling unnecessary. Maps of $\chi_\phi^2$ can be simply added together, and the resultant will weigh each constituent approximately equal, leading to a combined LC fit that ideally favours neither the radio nor the $\gamma$-ray fit.

\begin{figure}[t]
\centering
\includegraphics[scale=0.32]{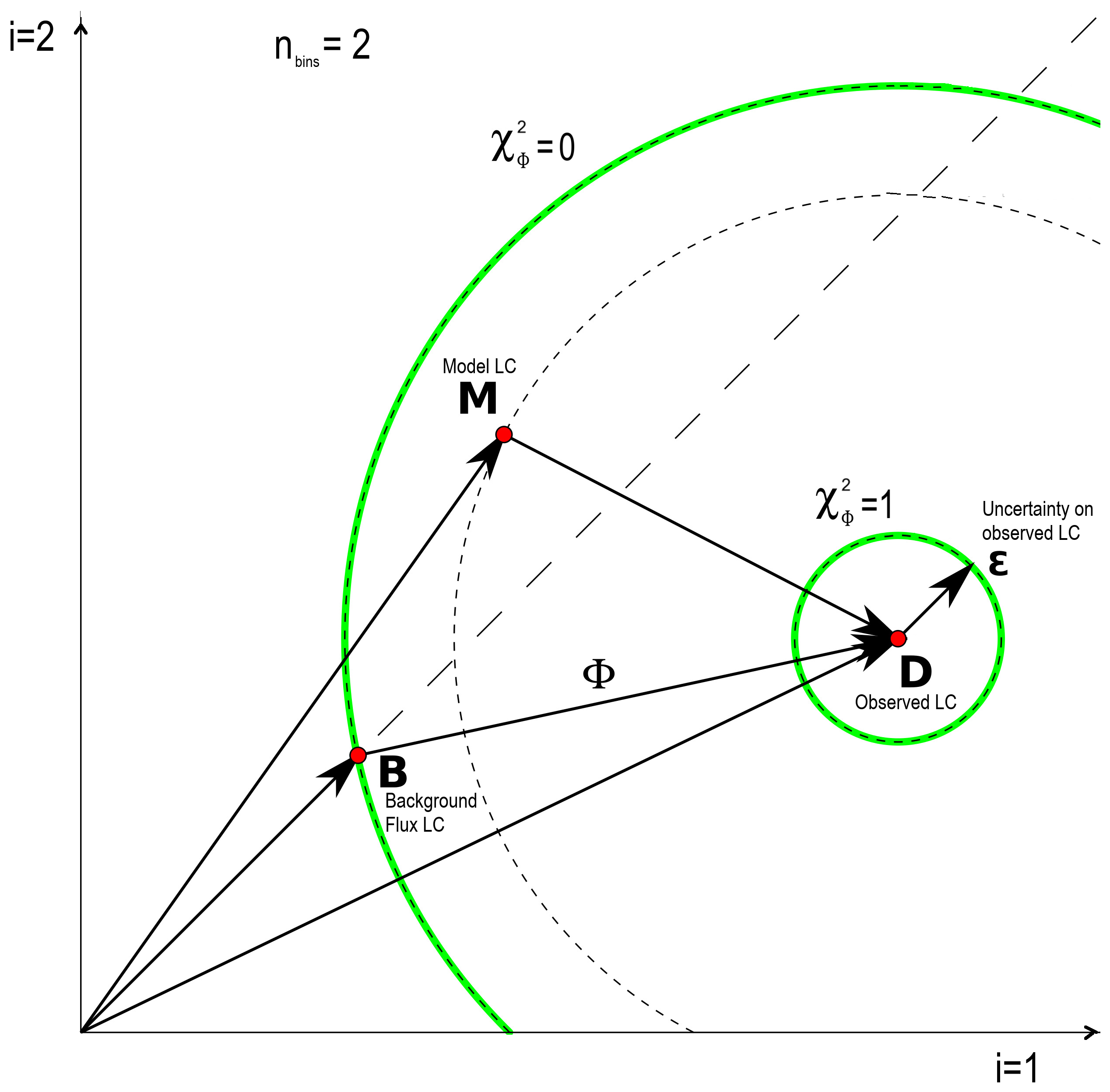}
\caption{\small{A two-dimensional representation of an $n_{\rm{bins}}$-dimensional space corresponding to $n_{\rm{bins}}$ phase bins of a pulsar LC. A representative model LC, observed LC, and background flux LC are represented by the points $M$, $D$, and $B$ in this space. Pearson's $\chi^2$ test statistic measures the distance between $M$ and $D$ in this space in units of observational errors, $\epsilon$, and the SFN test statistic converts this distance into units of the signal-to-noise ratio. The outer green circle centred on $D$ represents the zero-value of the $\chi^2_\phi$ test statistic, and the inner circle represents its unity value. Note that for this figure we assumed equal errors in both phase binds of the LC$-$in general each bin will have a slightly different relative error, and thus the inner boundary will be irregularly shaped. See the text for details.}}
\label{fig:heuristic}
\end{figure}

Figure \ref{fig:heuristic} presents a simple geometrical framework for thinking about test statistics that is illustrative of the differences between Pearson's $\chi^2$ test statistic and our SFN test statistic, and why the latter is better suited to multiband LC fitting.
One can envision an $n_{\rm{bins}}$-dimensional space where each dimension corresponds to one of $n_{\rm{bins}}$ phase bins into which a pulsar's LC is divided. The diagram in Figure \ref{fig:heuristic} is a two-dimensional analogue of this space. Any particular LC is represented by a single point in this space, e.g. a modelled LC $M$, an observed LC $D$, and a flat background LC $B$ as in the figure. Furthermore, we can construct vectors $\vec{M}$, $\vec{D}$, and $\vec{B}$ from the origin to these points. Goodness-of-fit tests can be thought of as concerned with measuring and comparing the length of and separation between such vectors in this space.

Pearson's $\chi^2$ test statistic measures the distance between $\vec{M}$ and $\vec{D}$ in this space in units of $\vec{\epsilon}$, the error on observations. It is valid to compare these "distances" from different model predictions to a single set of observations (say, a radio LC), but these distances are not necessarily comparable to distances from a second set of observations (say, a $\gamma$-ray LC). The difference in error values $\epsilon$ of various observations mean that $\chi^2$ distances in this space are measured in different units, and thus cannot be simply added.

The SFN test statistic scales the $\chi^2$ square of the distance between $\vec{M}$ and $\vec{D}$ to units of $\Phi$, the square of the distance between $\vec{B}$ and $\vec{D}$, thus normalising this distance. We can draw two concentric circles centred on $\vec{D}$, as in green on the figure, the inner circle being defined by $\vec{\epsilon}$, and the outer circle by $\vec{B}$. The closer $\vec{M}$ is to the outer circle, the closer $\chi^2_\phi$'s value is to zero; the closer $\vec{M}$ is to the inner circle, the closer $\chi^2_\phi$'s value is to one. This makes it clear that the SFN test statistic is significantly less dependent on observational uncertainties, and that distances measured in this way are more comparable between wavebands.

\section{Results}

In this section we relay the results we found by applying the LC fitting methods described in the previous section to a selection of 16 pulsars. We fit LCs obtained using all four possible combinations of the cone/core radio and TPC/OG geometrical $\gamma$-ray models to observed LCs from the Second \textit{Fermi} Pulsar Catalog \cite{abdo2013}. Table \ref{tab:constraints} collates the $(\alpha,\zeta)$ constraints we found using by-eye fitting, those we found using the SFN test statistic, and those found by the previous studies through manipulation of Pearson's $\chi^2$ for the same pulsars. The errors reported on the best-fit $(\alpha,\zeta)$ combinations produced by SFN fitting were obtained through Monte Carlo error estimation.

Figure \ref{fig:SFNvsBE_al} plots the best-fit estimates for the tilt angle $\alpha$ for each sample pulsar found through statistical fitting using the SFN test statistic against those found through by-eye fitting. Each colour represents a different combination of radio and $\gamma$-ray models used to obtain the best-fit $\alpha$. The closer a point is to the diagonal on this plot, the greater the agreement is between the fit we found by eye and that found using the SFN test statistic. In the legend of the figure, a value of Pearson's correlation coefficient, $r$, is assigned to every model combination. This is a measure of the correlation between the two variables compared in the plot. This coefficient takes on values between $-1$ and $+1$, where 1 indicates a total positive correlation, $-1$ a total negative correlation, and 0 no correlation. Table\ \ref{tab:correlations} compiles values of $r$ found when comparing the constraints found using each fitting method and each possible combination of models.

\begin{sidewaystable}
\tiny
\resizebox{23cm}{8.5cm}{
\begin{tabular}{ ||c||c c c c|c c c c|c c c c|| }
 \specialrule{.2em}{0em}{0em} 
  &  & \texttt{BY-EYE FITTING} & & &  & \texttt{INDEPENDENT SOURCES}& & & & \texttt{SCALED-FLUX NORMALISED TEST STATISTIC} & &\\ 
 \specialrule{.2em}{0em}{0em} 
 
 $\textbf{Pulsar}$ & $\textbf{OG + Cone}$ & $\textbf{OG + Core}$ & $\textbf{TPC + Cone}$ & $\textbf{TPC + Core}$ & $\textbf{Pierbattista OG}$ & $\textbf{Pierbattista TPC}$ & $\textbf{Johnson OG}$ & $\textbf{Johnson TPC}$ & \textbf{OG + Cone} & \textbf{OG + Core} & \textbf{TPC + Cone} & \textbf{TPC + Core}\\

 \specialrule{.2em}{.1em}{.1em} 
 \textbf{J0030+0451} & (60$\pm$8,78$\pm$4) & (75$\pm$5,70$\pm$4) & (75$\pm$7,59$\pm$6) & (75$\pm$5,57$\pm$8) & & & (88$\substack{+1 \\ -2}$,68$\pm$1) & (74$\pm$2,55$\substack{+3\\-1}$)  & (85$\pm$0,67$\pm$0) & (84$\substack{+0 \\ -1}$,70$\substack{+0 \\ -1}$) & (73$\pm$0,60$\pm$0) & (73$\pm$0,60$\pm$0)\\
            &                          &                          & (59$\pm$6,78 $\pm$ 5) & (55$\pm$ 6,77 $\pm$2) & & & & & & & & \\[0.1cm]
 \specialrule{.02em}{0em}{0em}

 \textbf{J0205+6449} & (65$\pm$5,88$\pm$2) & (86$\pm$3,80$\pm$3) & (86$\pm$4,60$\pm$7) & (85$\pm$5,79$\pm$4) & (73$\pm$2,90$\pm$2) & (75$\pm$2,86$\pm$2) & & & (74$\pm$0,85$\pm$0) & (84$\pm$0,86$\pm$0) & (74$\pm$0,85$\pm$0) & (84$\substack{+2 \\ -5}$,86$\substack{+1 \\ -3}$)\\
            &                          & (81$\pm$3, 87$\pm$3) & (67$\pm$4, 87$\pm$3) & (78$\pm$5,84$\pm$5) & & & & & & & & \\[0.1cm]
 \specialrule{.02em}{0em}{0em}

 \textbf{J0437-4715} & (31$\pm$10,62$\pm$4) & (58$\pm$6,40$\pm$7) & (57$\pm$9,26$\pm$10) & (52$\pm$7,32$\pm$10) & & & (76$\pm$1,46 $\pm$1) & (35$\pm$1,64$\pm$1) & (22$\pm$0,65$\pm$0) & (35$\substack{+5 \\ -1}$,61$\pm$1) & (54$\substack{+0 \\ -1}$,47$\substack{+0 \\ -2}$) & (34$\substack{+18 \\ -0}$,62$\substack{+1 \\ -13}$)\\
            & (63$\pm$7,32$\pm$8)  & (40$\pm$8,58$\pm$5) & (25$\pm$9,59$\pm$8)  & (27$\pm$10,51$\pm$8) & & & & & & & & \\[0.1cm]
 \specialrule{.02em}{0em}{0em}

 \textbf{J0631+1036} & (54$\pm$10,53$\pm$9) & (58$\pm$8,57$\pm$8) & (49$\pm$7,39$\pm$9) & (52$\pm$2,51$\pm$2) & (87$\pm$2,72$\pm$2) & (48$\pm$2,67$\pm$2) & & & (69$\substack{+8\\-4}$,61$\substack{+7\\-4}$) & (66$\substack{+2 \\ -3}$,65$\pm$2) & (62$\substack{+6 \\ -4}$,54$\substack{+6 \\ -4}$) & (60$\substack{+3 \\ -2}$,59$\substack{+4 \\ -1}$)\\
            &                           &                          & (36$\pm$5,49$\pm$6) &                         & & & & & & & &\\[0.1cm]
 \specialrule{.02em}{0em}{0em}

 \textbf{J0659+1414} & (57$\pm$3,49$\pm$2) & (52$\pm$3,52 $\pm$3) & (42$\pm$9,34$\pm$8) & (34$\pm$6,34$\pm$9) & (78$\pm$2,73$\pm$2) & (30$\pm$2,32$\pm$2) & &  & (54$\substack{+9\\-8}$,48$\substack{+14\\-3}$) & (56$\substack{+0 \\ -3}$,52$\substack{+4 \\ -0}$) & (53$\substack{+1 \\ -0}$,46$\substack{+0 \\ -1}$) & (51$\pm$2,51$\substack{+2 \\ -3}$)\\
            & (48$\pm$2,58$\pm$3) &                          & (34$\pm$6,43$\pm$6) &                         & & & & & & & & \\[0.1cm]
 \specialrule{.02em}{0em}{0em}

 \textbf{J0742-2822} & (77$\pm$4,64$\pm$4) & (82$\pm$7,81$\pm$7) & (48$\pm$5,38$\pm$5) & (40$\pm$13,39$\pm$11) & (76$\substack{+2\\-3}$,86$\substack{+3\\-2}$) & (62$\pm$2,77$\pm$2) & & & (77$\substack{+1\\-10}$,88$\substack{+1\\-7}$) & (85$\pm$3,85$\substack{+2 \\ -3}$) & (66$\substack{+8 \\ -7}$,79$\substack{+6 \\ -2}$) & (83$\substack{+2 \\ -3}$,83$\substack{+2 \\ -3}$)\\
            & (74$\pm$4,85$\pm$4) & (49$\pm$3,49$\pm$3) & (38$\pm$2,50$\pm$2) &                           & & & & & & & & \\[0.1cm]
 \specialrule{.02em}{0em}{0em}

 \textbf{J1124-5916} & (56$\pm$8,86$\pm$3) & (83$\pm$6,84$\pm$5) & (84$\pm$6,62$\pm$6) & (82$\pm$8,82$\pm$8) & (83$\pm$2,88$\pm$2) & (84$\pm$2,89$\pm$2) & & & (71,88) & (89,83) & (83,80) & (83,82)\\
            &                          &                          & (67$\pm$8,84 $\pm$5) &                         & & & & & & & &\\[0.1cm]
 \specialrule{.02em}{0em}{0em}

 \textbf{J1231-1411} & (79$\pm$5,50$\pm$4) & (81$\pm$8,53$\pm$10) & (75$\pm$5,46$\pm$7) & (73$\pm$5,48$\pm$9) & & & (88$\pm$1,67$\pm$1) & (26$\substack{+3\\-4}$,69 $\pm$ 1) & (82$\pm$0,65$\pm$0) & (80$\pm$0,66$\substack{+1 \\ -0}$) & (71$\pm$0,59$\pm$0) & (71$\pm$0,59$\pm$0)\\
            & (85$\pm$2,59$\pm$3) & (55$\pm$3,77$\pm$2)  & (46$\pm$6,76$\pm$2) & (47$\pm$5,76$\pm$2) & & & & & & & &\\[0.1cm]
 \specialrule{.02em}{0em}{0em}

 \textbf{J1410-6132} & (63$\pm$4,40$\pm$5) & (59$\pm$3,52$\pm$4) & (53$\pm$9,35$\pm$9) & (37$\pm$8,44$\pm$9) & (87$\pm$2,76$\pm$2) & (19$\substack{+2 \\ -4}$,6 $\pm$ 2) & & & (52$\substack{+37\\-17}$,84$\substack{+5\\-51}$) & (83$\substack{+6 \\ -34}$,89$\substack{+0 \\ -40}$) & (6$\substack{+9 \\ -5}$,17$\substack{+12 \\ -3}$) & (5$\substack{+10 \\ -3}$, 10$\substack{+5 \\ -0}$)\\
                     & (11$\pm$9,12$\pm$9) & (7$\pm$4,6$\pm$4)   & (4$\pm$3, 20$\pm$3)  &                         & & & & & & & & \\[0.1cm]
 \specialrule{.02em}{0em}{0em}

 \textbf{J1420-6048} & (60$\pm$6,61$\pm$6) & (61$\pm$2,60$\pm$2) & (54$\pm$2,46$\pm$2) & (50$\pm$1,50$\pm$1) & (55$\pm$2,57$\pm$2) & (52$\pm$2,53$\pm$2) & &  & (65$\substack{+1\\-0}$,50$\pm$1) & (60$\substack{+1 \\ -0}$,58$\pm$1) & (60$\pm$0,46$\pm$0) & (54$\substack{+2 \\ -1}$,53$\substack{+2 \\ -1}$)\\[0.1cm]
 \specialrule{.02em}{0em}{0em}

 \textbf{J1509-5850} & (58$\pm$5,59$\pm$6) & (54$\pm$4,55$\pm$5) & (54$\pm$2,52$\pm$2) & (54$\pm$3,54$\pm$5) & (85$\pm$2,76$\pm$2) & (46$\pm$2,66$\pm$2) & & & (76$\substack{+7\\-14}$,65$\substack{+6\\-14}$) & (67$\substack{+3 \\ -11}$,66$\substack{+2 \\ -10}$) & (48$\substack{+3 \\ -1}$,60$\pm$2) & (51$\substack{+2 \\ -1}$,53$\substack{+1 \\ -3}$)\\[0.1cm]
 \specialrule{.02em}{0em}{0em}

 \textbf{J1513-5908} & (58$\pm$4,42$\pm$4) & (52$\pm$4,52$\pm$4) & (48$\pm$8,33$\pm$7) & (33$\pm$8,38$\pm$8) & (60$\pm$2,59$\pm$2) & (50$\pm$2, 54$\pm$2) & & & (56$\substack{+1\\-2}$,47$\pm$1) & (52$\substack{+1 \\ -2}$,51$\substack{+2 \\ -1}$) & (55$\pm$1,46$\substack{+2 \\ -1}$) & (52$\substack{+0 \\ -1}$,51$\substack{+1 \\ -0}$)\\
                     & (41$\pm$4,59$\pm$4) &                          & (35$\pm$5,51$\pm$7) &   & & & & & & & & \\[0.1cm]
 \specialrule{.02em}{0em}{0em}

 \textbf{J1614-2230} & (54$\pm$6,84$\pm$4) & (52$\pm$8,83$\pm$6) & (85$\pm$5,54$\pm$7) & (84$\pm$5,51$\pm$6) & & & (64$\substack{+8\\-20}$,88$\substack{+2\\-5}$) & (80$\substack{+8\\-20}$,80$\substack{+6\\-4}$) & (89$\substack{+0\\-5}$,84$\substack{+2\\-3}$) & (88$\substack{+1 \\ -3}$,84$\substack{+2 \\ -4}$) & (84$\substack{+2 \\ -1}$,82$\pm$1) & (85$\substack{+1 \\ -6}$,83$\substack{+3 \\ -1}$)\\
                     &                          &                          & (55$\pm$6,84$\pm$5) & (54$\pm$2,81$\pm$2) & & & & & & & & \\[0.1cm]
 \specialrule{.02em}{0em}{0em}

 \textbf{J1718-3825} & (51$\pm$4,51$\pm$4) & (54$\pm$4,53$\pm$4) & (46$\pm$7,34$\pm$8) & (34$\pm$7,40$\pm$7) & (80$\pm$2,55$\pm$2) & (45$\pm$2,65$\pm$2) & & & (67$\substack{+5\\-2}$,56$\substack{+2\\-5}$) & (67$\pm$3,63$\pm$3) & (47$\pm$0,54$\pm$0) & (57$\pm$0,58$\pm$0)\\
                     &                          &                          & (32$\pm$8,47$\pm$8) &                          & & & & & & & & \\[0.1cm]
 \specialrule{.02em}{0em}{0em}

 \textbf{J1833-1034} & (55$\pm$10,78 $\pm$3) & (78$\pm$5,72$\pm$5) & (75$\pm$5,52$\pm$8) & (53$\pm$8,78$\pm$2) & (65$\pm$2,87$\pm$2) & (55$\pm$2,75$\pm$2) & & & (87$\substack{+2\\-6}$,65$\substack{+2\\-6}$) & (85$\substack{+3 \\ -1}$,81$\pm$2) & (67$\substack{+0 \\ -1}$,80$\substack{+0 \\ -1}$) & (80$\substack{+6 \\ -0}$,83$\substack{+3 \\ -2}$)\\
                     & (85$\pm$5,64$\pm$7)  & (71$\pm$5,77$\pm$4) & (53$\pm$8,78$\pm$2) &                          & & & & & & & & \\[0.1cm]
 \specialrule{.02em}{0em}{0em}

 \textbf{J2229+6114} & (60$\pm$5,38$\pm$6) & (50$\pm$5,50$\pm$5) & (51$\pm$10,33$\pm$8) & (34$\pm$6,41$\pm$7) & (75$\pm$2,55$\pm$2) & (42$\pm$2,55$\pm$2) & & & (64$\pm$1,50$\pm$1) & (61$\substack{+0 \\ -2}$,59$\pm$2) & (51$\substack{+0 \\ -3}$,65$\substack{+1 \\ -3}$) & (54$\pm$0,45$\pm$0)\\
                     &                          &                          & (30$\pm$9,52$\pm$8)  &                          & & & & & & & & \\[0.1cm]   
 \specialrule{.2em}{.1em}{.1em} 
\end{tabular}
}
\caption{\small{Best-fit ($\alpha$,$\zeta$) pairs (in degrees) for each sample pulsar using by-eye fitting, those found by other authors, and those we found using the SFN test statistic. We used four different combinations of the OG and TPC $\gamma$-ray models and the core and cone radio models for fitting.}}
\label{tab:constraints}
\end{sidewaystable}

\section{Discussion}
\begin{figure}[b!]
\centering
\resizebox{15cm}{8cm}{
\includegraphics[width=\textwidth]{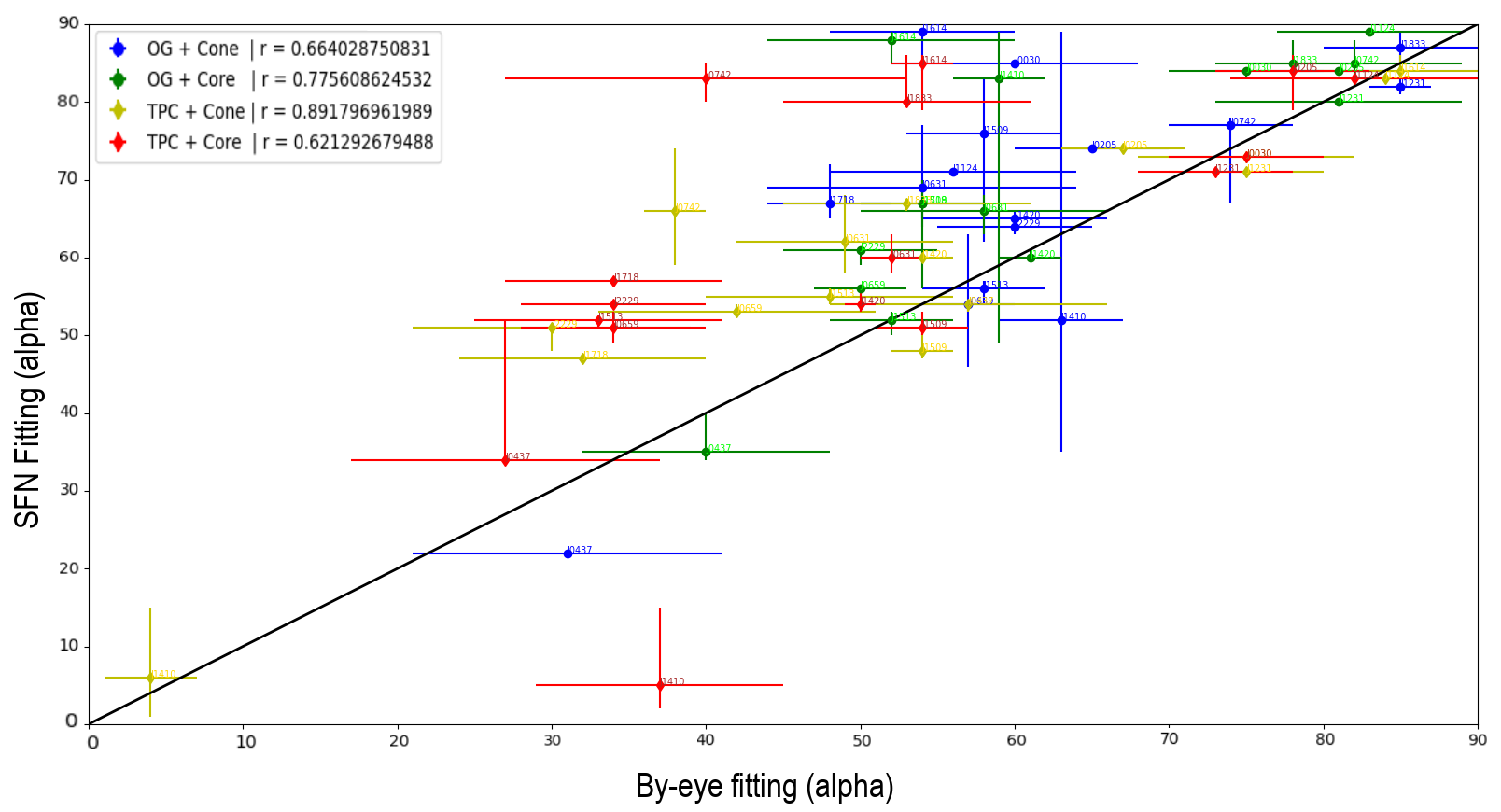}
}
\caption{\small{Comparison of best-fit tilt angles ($\alpha$) found using the SFN test statistic to those found by eye. Each colour represents constraints found by employing a different combination of radio and $\gamma$-ray models.}}
\label{fig:SFNvsBE_al}
\end{figure}

It is apparent from Figure \ref{fig:SFNvsBE_al} that although there is general agreement between the $\alpha$ constraints we found using SFN fitting and those we found by eye, this correlation is fairly sensitive to the combination of radio and $\gamma$-ray models employed. In particular, the combinations of the OG + Cone models (blue) and TPC + Core models (red) produce a large number of outliers and significantly lower values of $r$ than is the case for the OG + Core (green) and TPC + Cone (yellow) combinations. The former pair of model combinations also seem to produce fits with generally larger error bars$-$perhaps indicative of a failure of these models to produce any one particularly good candidate LC. It is important to bear in mind our the purpose for by-eye fitting as a "sanity check" to ensure that statistical fitting methods are not producing qualitatively incredible fits. We would suggest that the fact of broad accord between our SFN fits and by-eye fits supports the case of the SFN fitting method as useful, at least in the sense that it does not demonstrate the opposite.

\begin{table}[t]
\centering
\resizebox{15cm}{1.2cm}{
\begin{tabular}{|l||ll|ll|ll|ll|}
 \specialrule{.2em}{0em}{0em}
  & \texttt{\small{OG + Cone}} & & \texttt{\small{OG + Core}} & & \texttt{\small{TPC + Cone}} & & \texttt{\small{TPC + Core}} &\\
  &\small{$\alpha$}&\small{$\zeta$}&\small{$\alpha$}&\small{$\zeta$}&\small{$\alpha$}&\small{$\zeta$}&\small{$\alpha$}&\small{$\zeta$}\\
 \specialrule{.2em}{0em}{0em} 
\texttt{\small{SFN vs by eye}} & \small{0.6640} & \small{0.7167} & \small{0.7756} & \small{0.7398} & \small{0.8918} & \small{0.8454} & \small{0.6213} & \small{0.6562}\\
\texttt{\small{Other authors vs by-eye}} & \small{-0.0102} & \small{0.6371} & \small{0.1777} & \small{0.7380} & \small{0.6639} & \small{0.7623} & \small{0.5895} & \small{0.6299}\\
\texttt{\small{SFN vs other authors}} & \small{0.0108} & \small{0.7308} & \small{0.2143} & \small{0.7843} & \small{0.7629} & \small{0.8913} & \small{0.7803} & \small{0.9156}\\
\specialrule{.2em}{0em}{0em}
\end{tabular}
}
\caption{\small{Values of the Pearson's correlation coefficient $r$ for the $\alpha$ and $\zeta$ best-fit estimations found using all different fitting methods and model combinations.}}
\label{tab:correlations}
\end{table}
 
As is evident from Table \ref{tab:correlations}, the fits found using Pearson's $\chi^2$ test do not seem to correlate with the by-eye fits as well as the SFN fits do for either $\alpha$ or $\zeta$. While the lower correlation between the by-eye fits and these authors' fits does not necessarily establish the inadequacy of the fitting methods they used, it does suggest that SFN fitting produces superficially better LC fits. Table \ref{tab:correlations} also demonstrates a general agreement between the constraints found using Pearson's $\chi^2$ test and our SFN fits for both parameters, besides two outliers, namely the $\alpha$ constraints found using the OG + Cone and the OG + Core combinations of models. It is telling to view these poor correlations in the light of the similar lack of agreement between these $\alpha$ fits found by the other authors and those we found by eye. This fact leads us to conclude that the previous authors' $\alpha$ constraints found using the OG $\gamma$-ray model might not be entirely trustworthy, and that their disaccord with what we found through SFN fitting is not necessarily evidence of a problem with this fitting method. Apart from these two outliers, however, there seems to be a strong correlation overall between the parameter constraints we found using the SFN test statistic and those found by using Pearson's $\chi^2$ test statistic.

Taken together, Tables \ref{tab:constraints} and \ref{tab:correlations} suggest that SFN pulsar LC fitting delivers best-fit LCs that are more convincing to the eye$-$in both the radio and $\gamma$-ray bands simultaneously$-$than those produced by the manipulation of Pearson's $\chi^2$ test statistic as put into practice in previous studies.

\section{Conclusions}
Our aim was to test the utility of the novel SFN test statistic put forward by Seyffert et al. (2018; in preparation) as it applies to dualband pulsar LC fitting. We believe, based on the correlations laid out in Table \ref{tab:correlations}, that the use of this test statistic results in LC fits that are more aligned to what one typically finds through by-eye fitting than is the case with fits found through manipulation of Pearson's $\chi^2$ test statistic. As such, fitting using the SFN test statistic may be useful for any further pulsar LC modelling efforts.

Future work may focus on applying this fitting method to LCs modelled using a variety of different radio and $\gamma$-ray models, including the new dissipative $\gamma$-ray models (e.g. \cite{kalap2014}), in order to assess the merit of these models. This fitting procedure may also be extended to incorporate different wavebands such as x-rays, or even to multiband fitting of pulsars' spectra.

\end{document}